\documentclass[12pt]{article}
\usepackage{epsfig,amsfonts,amssymb}
\usepackage{hyperref}
\pdfoutput=1

\usepackage{cite}
\usepackage{comment}

\def\ZZZ{{\hbox{ Z\kern-1.6mm Z}}}
\def\RRR{{\hbox{ R\kern-2.4mm R}}}
\def\CCC{{\hbox{ C\kern-2.0mm C}}}
\def\zzz{{\hbox{z\kern-1mm z}}}

\def\ZZZ{\mathbb{Z}}
\def\RRR{\mathbb{R}}

\newcommand{\qeq}{{\hbox{=\kern-2.3mm ? \kern.5mm }}}
\renewcommand{\qeq}{=}

\newcommand{\II}{{\cal I}}

\newcommand{\NN}{{\cal N}}

\newcommand{\be}{\begin{equation}}
\newcommand{\ee}{\end{equation}}
\newcommand{\ben}{\begin{eqnarray}\displaystyle}
\newcommand{\een}{\end{eqnarray}}

\newcommand{\refb}[1]{(\ref{#1})}
\newcommand{\p}{\partial}
\newcommand{\sectiono}[1]{\section{#1}\setcounter{equation}{0}}

\def\one{{\hbox{ 1\kern-.8mm l}}}
\def\zero{{\hbox{ 0\kern-1.5mm 0}}}

\newcommand{\bea}[1]{\begin{eqnarray}\label{#1} }
\newcommand{\eea}{\end{eqnarray}}

\newcommand{\eqref}{\refb}

\usepackage{bm}
\usepackage[table]{xcolor}

\begin{document}

\baselineskip 24pt

\begin{center}

{\Large \bf Logarithm of charge ratio in black hole entropy}

\end{center}

\vskip .6cm
\medskip

\vspace*{4.0ex}

\baselineskip=18pt

\centerline{\large \rm Muktajyoti Saha, Ashoke Sen,  P. Shanmugapriya}

\vspace*{4.0ex}

\centerline{\large \it International Centre for Theoretical Sciences - TIFR 
}
\centerline{\large \it  Bengaluru - 560089, India}


\vspace*{1.0ex}
\centerline{\small E-mail:  muktajyoti.saha@icts.res.in, ashoke.sen@icts.res.in, 
shanmugapriya.prakasam@icts.res.in}

\vspace*{5.0ex}

\centerline{\bf Abstract} \bigskip

Logarithmic correction to BPS black hole entropy, computed from microscopic description, often
contains terms involving large ratios of charges, besides the logarithmic terms involving the overall
scale of the charges. If the electric charges are much larger than the magnetic charges,
then the attractor value of the string coupling is small and one might hope to use weakly coupled string
theory to compute logarithmic corrections involving ratios of charges from the macroscopic
side. We compute these for black holes in flat space-time, preserving four supercharges,
in $\mathcal{N} = 2$, $\mathcal{N}=4$ and $\mathcal{N}=8$ supersymmetric string compactifications
in four dimensions. We find perfect agreement with the microscopic results in $\mathcal{N}=4$ and 
$\mathcal{N}=8$ theories, for which the microscopic results are known. Various stringy and statistical mechanical 
effects become important in this analysis, including 1) use of the correct ultra-violet cut-off (string scale
instead of Planck scale),  2) correct path integral measure (ultra-local measure with appropriate 
dilaton dependent metric), 3) use of the correct path integral variable
(Kalb-Ramond 2-form instead of the dual axion) and 4) change of ensemble (from grand canonical to
microcanonical). We also verify that the measure we use is consistent with what
follows from the BV formalism of string field theory.

\vfill \eject

\tableofcontents

\sectiono{Introduction} \label{s0}

Black hole entropy in a quantum theory of gravity has a universal term given by the Bekenstein-Hawking
formula. It also has a semi-universal term proportional to the logarithm of the area that depends only on
the spectrum of massless states and their interaction in the low energy effective field theory, but is
insensitive to the spectrum and interaction of massive states or the higher derivative terms in the low
energy effective field theory.  For many supersymmetric black holes the leading Bekenstein-Hawking
entropy and the subleading logarithmic corrections have been shown to agree with the result of
microscopic counting of the black hole states.\footnote{Throughout this paper we shall
define entropy as the logarithm of the appropriate helicity supertrace index to be defined later.}

The goal of this paper will be to explore the finer details of the logarithmic correction to the black hole
entropy. Earlier analysis of logarithmic corrections was performed assuming that the near horizon geometry
of the black hole has a single large parameter, -- the area of the event horizon. In particular the string coupling
was assumed to be of order unity so that the string scale and the Planck scale are of the same order, and one
did not need to worry about whether the argument of the log involves the horizon area measured in string scale or
the Planck scale. However for certain supersymmetric black holes, by taking the electric charges carried by
the black hole to be much larger than the magnetic charges, 
one can ensure that the near horizon value of the
string coupling is small. In this case we have two large near horizon parameters, the horizon area measured
in the string scale and the inverse of the near horizon value of the string coupling. In terms of the charges, the large
parameters can be taken to be the overall scale of the magnetic charge and the ratio of the electric and the
magnetic charges. Consequently the black hole entropy can have two types of logarithmic terms: the
logarithm of the magnetic charge and the logarithm of the ratio of the electric and the magnetic charge.

Since the near horizon string coupling is weak, one might expect that we can use
perturbative string theory to compute the coefficients of these two type of logarithmic terms separately.
This is the task we undertake in this paper. In particular, we compute this for black holes preserving four
supercharges in a wide class of heterotic
and type II string compactifications to 3+1 dimensional Minkowski space with $\NN=4$ and $\NN=8$
supersymmetry using the procedure for computing black hole index described in 
\cite{Cabo-Bizet:2018ehj,Iliesiu:2021are,H:2023qko,Anupam:2023yns}. 
For these black holes the exact result of microscopic counting is known. From these one can 
extract the coefficients of the logarithmic terms and we find perfect agreement between the microscopic and the
macroscopic results in all cases.

We now summarize the main results. Let us denote by $\lambda_Q$ and $\lambda_P$ the scales of the electric
and magnetic charges carried by the black hole. Then we find that both the macroscopic and the microscopic
analysis give the following logarithmic correction to the black hole
entropy:
\ben\label{e2pre}
\hbox{$\NN=4$ theories} &:&  \qquad -{r+2\over 2} \, \ln \, {\lambda_Q\over \lambda_P}\, , \nonumber \\
\hbox{$\NN=8$ theory} &:& \qquad -8 \, \ln\, \lambda_P -4\,  \ln \, {\lambda_Q\over \lambda_P} \, ,
\een
where, for the $\NN=4$ theories, $r$ denotes the total number of matter multiplets in the theory. For the
$\NN=8$ theory, the spectrum and interaction are completely fixed by supersymmetry in the low energy
limit.

We also derive similar results for $\NN=2$ supersymmetric heterotic string theory in four dimensions
with the result:
\be
{1\over 6} \, (23 - n_V + n_H)
\, \ln \lambda_P +  \left({3 - n_V \over 2}
\right)\, \ln{\lambda_Q\over \lambda_P}\, ,
\ee
where $n_V$ and $n_H$ are, respectively, the number of vector multiplets and hypermultiplets.
In this case there are no known microscopic counting results that one can compare this with.

During this analysis we encounter various subtle points that we list below:
\begin{enumerate}
\item We use the method of \cite{Cabo-Bizet:2018ehj,Iliesiu:2021are,H:2023qko,Anupam:2023yns} 
to compute the index of the black holes under consideration. This
gives the result in the grand canonical ensemble. While extracting the microcanonical entropy from this by
Fourier transformation, we encounter additional logarithms of the ratios of electric and magnetic charges
that need to be accounted for.
\item String theory contains a Kalb-Ramond 2-form field which is often dualized to a scalar field while writing the
low energy effective action. However the result of path integral over the scalar and the 2-form field differ
by contribution from the zero modes of the two form field. These generate logarithms of  the 
ratios of electric and magnetic charges
that need to be accounted for.
\item Since for small string coupling the string scale and the Planck scale differ by a large number,
it makes a difference whether in computing the logarithmic corrections due to the non-zero modes
of the fields we take the ultra-violet cut-off to
be string scale or the Planck scale. In a weakly coupled string theory the natural ultra-violet cut-off is the
string scale. We use this for computing the contribution from the non-zero modes.
\item The choice of UV cut-off 
is also related to the integration measure that we need to use while carrying out path integral over
the zero modes of various fields. This is turn can be determined from the BV master action of string field theory. 
We use
this measure for carrying out the path integral over the fields and show that this is
also compatible with using the
string scale as the UV cut-off. 
\end{enumerate}

\sectiono{The problem} \label{s1}

We shall analyze supersymmetric black holes, carrying four unbroken supersymmetries,  in 
heterotic string theory on $T^6\times R^{3,1}$\cite{Dijkgraaf:1996it,LopesCardoso:2004law}, 
the 3+1 dimensional CHL type models\cite{Chaudhuri:1995fk} with $\NN=4$ supersymmetry and
type II string theory on $T^6\times R^{3,1}$\cite{Shih:2005qf,Pioline:2005vi}.  
The CHL type models that we consider include all
models analyzed 
in \cite{Jatkar:2005bh,David:2006yn,David:2006ru,David:2006ud,Sen:2007qy}. 
In these models
we start from type II or heterotic string theory on $T^6$ and take
a quotient by a transformation 
that acts on only the left-moving world-sheet fields, together with a shift along one of the circles. 
This projects out all the RR sector massless fields in the type II theory and 
preserves all supersymmetries
arising from the right-moving sector of the world-sheet. 
We shall refer to type II on $T^6\times R^{3,1}$ as the $\NN=8$ theory and
the heterotic string theory on $T^6\times R^{3,1}$, as well as all the CHL type models,
collectively as $\NN=4$ theories.
The black holes in the $\NN=8$ theory could carry both NSNS and RR sector charges, but 
we shall  restrict the charge vector to lie in the NSNS sector.
In this case, in all of these theories, 
the leading Bekenstein Hawking entropy
for all of these black holes have the form
\be\label{eclassical}
S_{BH} = \pi\sqrt{Q^2 P^2 - (Q.P)^2} \, ,
\ee
where $Q$ and $P$ are vectors of electric and magnetic charges carried by the black hole, and
\be
Q^2=Q^T L Q, \qquad P^2 = P^T L P, \qquad Q.P = Q^T L P\, ,
\ee
where $L$ is a matrix with 6 eigenvalues $-1$,  6 eigenvalues $+1$ for the $\NN=8$ theory and $r$ 
eigenvalues +1 for the $\NN=4$ theories. The value of $r$ depends on the particular model we are
considering.
For $T^6$ compactification of heterotic string theory we have $r=22$.

Logarithmic corrections to the entropy have been computed from  path integral over massless
fields in the near horizon $AdS_2\times S^2$ 
geometry\cite{Banerjee:2010qc,Banerjee:2011jp,Keeler:2014bra} 
as well as the full geometry\cite{H:2023qko,Anupam:2023yns}. 
The computation was done in the limit when
$Q$ and $P$ are both large. In this limit the size of the black hole scales as the charge and a logarithm
of the size $a$ of the black hole can be translated to a logarithm of the charge. Computations based
on the near horizon $AdS_2\times S^2$ geometry and the full geometry agree, and the result 
for the logarithmic correction is:
\ben\label{e1}
\hbox{$\NN=4$ theories} &:&  \qquad 0\, , \nonumber \\
\hbox{$\NN=8$ theory} &:& \qquad -8 \, \ln\, a \, .
\een
Both the leading contribution \refb{eclassical} and the logarithmic correction \refb{e1} agree
with the results of microscopic counting\cite{Dijkgraaf:1996it,LopesCardoso:2004law,Jatkar:2005bh,David:2006yn,
Sen:2007qy,Shih:2005qf,Pioline:2005vi,Sen:2009gy,Banerjee:2010qc,Banerjee:2011jp}.

In this paper we shall consider a situation where the electric charges are much larger than the
magnetic charges even though both charges are large. Let us denote by $\lambda_Q$ and 
$\lambda_P$ the scales of the electric and the magnetic charges. In this case we expect
that there should also be corrections to the entropy proportional to $\ln(\lambda_Q/\lambda_P)$.
Let us parametrize the net logarithmic corrections to the entropy in these two theories as
\ben\label{e2}
\hbox{$\NN=4$ theories} &:&  \qquad c_H \, \ln \, {\lambda_Q\over \lambda_P}\, , \nonumber \\
\hbox{$\NN=8$ theory} &:& \qquad -8 \, \ln\, \lambda_P + c_{II}  \ln \, {\lambda_Q\over \lambda_P} \, ,
\een
so as to be consistent with \refb{e1} when both $\lambda_P$ and $\lambda_Q$ are of order $a$.
Our goal will be to compute the coefficients $c_H$ and $c_{II}$.

The microscopic results for these ratios are known. 
In the $\NN=4$ theories, the 
result  is\cite{Dijkgraaf:1996it,LopesCardoso:2004law,Jatkar:2005bh,David:2006yn,
Sen:2007qy}\footnote{This can be seen, e.g.\ from equation (3.1.25), (3.1.40) and (3.1.49) 
of \cite{Sen:2007qy} after
accounting for the fact that $r$ in this paper is $r-6$ of \cite{Sen:2007qy}.
For the 
heterotic string
compactifications there is also a term linear in $\lambda_Q$, but this can be 
attributed to string tree level four derivative correction to the effective action. See e.g. 
eqs.(3.1.50)-(3.1.53) of \cite{Sen:2007qy} for a review.}
\be\label{e3}
c_H = -{r+2\over 2} \, .
\ee
On the other hand in $\NN=8$ theory, the full U-duality group forces the entropy to be a
function of $\sqrt{Q^2P^2-(Q.P)^2}$, i.e. of the combination $\lambda_Q\lambda_P$.
Hence it follows from \refb{e2} that
\be \label{e4}
c_{II}=-4\, .
\ee
This also agrees with the result of microscopic counting\cite{Sen:2009gy}.

A different approach to computing the coefficients $c_H$ from the macroscopic side was
adopted in \cite{LopesCardoso:2004law,Jatkar:2005bh,David:2006yn}, where the
four derivative correction to the effective action was first computed using a dual description
of the theory following \cite{Gregori:1997hi} and then that action was used to compute the correction to the
entropy using Wald's general formula\cite{Wald:1993nt}. 
Our goal will be to compute the entropy directly in the
description where the string coupling is small instead of relying on any conjectured duality
symmetry.

For later use, it will be useful to review some features of the near horizon geometry of the
black hole. 
From \refb{e1} it follows that in the Einstein frame, the size of the black hole is of order
\be\label{e5}
a_E \sim \{Q^2P^2 -(Q.P)^2\}^{1/4}\sim \sqrt{\lambda_Q\lambda_P}\, .
\ee
On the other hand the near horizon value of the four dimensional dilaton $\phi$ is 
given by (see {\it e.g.} \cite{Sen:2007qy})
\be\label{e6}
e^{-2\phi} = { \sqrt{Q^2P^2 -(Q.P)^2}\over P^2 } \sim {\lambda_Q\over \lambda_P}\, .
\ee
Hence the size of the horizon, measured in the string metric, takes the form
\be\label{e7}
a_S \sim e^{\phi} a_E \sim \lambda_P\, .
\ee

Since we work in the limit $\lambda_Q>>\lambda_P$, we see from \refb{e6} that the string coupling 
$e^\phi$ at the horizon is small in this limit. Hence we can use perturbative string theory for our 
analysis. In this case the natural ultra-violet (UV) cut-off is the string scale. On the other hand the
natural infrared (IR) cut-off  is the horizon size, which, when measured in the string scale, is
$a_S\sim \lambda_P$. Hence $a_S$ also represents the ratio of the IR and UV cut-off.
In our analysis we shall express the arguments of the logarithmic terms in terms of $a_S\sim
\lambda_P$ and
the near horizon value of the string coupling
\be\label{e8}
g_s = e^\phi\sim \sqrt{\lambda_P\over \lambda_Q}\, .
\ee

\sectiono{Computation of the logarithmic terms} \label{s2}

In this section we shall compute the logarithmic terms in $\NN=4$ supersymmetric
compactifications of heterotic and type II string theories and $\NN=8$ supersymmetric compactification
of type II string theory. For this we shall use the macroscopic expression for the index based on the
full asymptotically flat space-time geometry, following 
\cite{Cabo-Bizet:2018ehj,Iliesiu:2021are,H:2023qko,Anupam:2023yns}. The advantage of using this
formalism is that one does not need to worry about possible logarithmic contribution from the hair modes,
-- the modes living outside the horizon.\footnote{In particular while the hair modes described in 
\cite{Banerjee:2009uk,Jatkar:2009yd}
do not give corrections of order $\ln \sqrt{\lambda_Q\lambda_P}$, they do give 
contributions proportional to $\ln (\lambda_Q/\lambda_P)$ for $\lambda_Q>>\lambda_P$.}

We begin by reviewing the procedure described in 
\cite{Cabo-Bizet:2018ehj,Iliesiu:2021are,H:2023qko,Anupam:2023yns} for computing the black hole
index. 
If the black hole of interest breaks $2n$ space-time supersymmetries, then the relevant 
index is\cite{Bachas:1996bp,Gregori:1997hi}\footnote{With this choice of normalization, $B_n$ counts
the number of BPS supermultiplets with sign, the sign being + ($-$) if the maximum helicity of the
states in the supermultiplet is integer (integer + half). \label{fo4}} 
\be 
B_n(P,Q) = {(-1)^{n/2}\over n!}\, Tr_{P,Q,\vec k=0}[(-1)^F e^{-\beta H} (2h)^n]\, ,
\ee
where $Tr_{P,Q,\vec k=0}$ denotes that the trace is taken over all states at rest, 
carrying fixed charges
$P,Q$ and $h$ is the third component of the angular momentum in the rest frame. Rotational 
invariance can be used to show that the result does not depend on the choice of the third axis.
The role of the $(2h)^n$ factor in the trace is to prevent the cancellation between the bosonic and
fermionic state contribution that arises due to $2n$ broken supersymmetry generators acting on a state,
creating new state with opposite statistics. The index can be shown to 
receive contribution only from supersymmetric states that break at most $2n$ supersymmetries.
All of these states have the same mass $M_{BPS}(P,Q)$.
If we denote by $e^{S_{BPS}(P,Q)}$ the number of such states, counted with sign as described
in footnote \ref{fo4}, then up to an overall sign, we have
\be
B_n( P, Q) = e^{S_{BPS}(P,Q)-\beta M_{BPS}(P,Q)}\, .
\ee

The strategy for computing $B_n$ will be to begin with the Euclidean path integral over all fields
with appropriate boundary conditions set on the fields at $\infty$. These boundary conditions
are set by the magnetic charges $P_i$, the inverse temperature $\beta$, the chemical potential
$\mu_i$ dual to the electric charges $Q_i$ and the angular velocity $\omega$, which can also
be interpreted as the chemical potential dual to the third component of the angular 
momentum\cite{Gibbons:1976ue}. $\beta$ determines the period of the Euclidean time $\tau$,
$\mu_i$'s determine the asymptotic values of the time component of the gauge fields,
and $\beta\omega$ determines the shift in the azimuthal angle $\phi$ that must accompany the
$\tau\to \tau+\beta$ transformation for periodic identification of the coordinates.
We also insert a factor of $(2J_3)^n$ in the path integral, where $J_3$ is the third
component of the angular momentum,  constructed
from the fields by the usual Noether procedure. Thus we can write:
\be
Z_n(\beta, P,\mu,\omega) = \int_{\rm Fields} e^\II \, (2J_3)^n\, ,
\ee
$\II$ being the euclidean action, related to the Lorentzian action $\II_L$ as $\II=i\II_L$. 
Such a path integral computes\cite{Gibbons:1976ue}
\be
Z_n(\beta, P,\mu,\omega) = Tr_P[e^{-\beta H + \beta\mu.Q + \beta \omega h} (2 h)^n]\, ,
\ee
where $Tr_P$ denotes trace over all states carrying fixed magnetic charges $P_i$.
It should be understood that in this formula $P,Q,\mu$ are multi-component variables since
there are multiple gauge fields in the theory.
For the special case of $\beta\omega=2\pi i$, we get, using $e^{2\pi i h}=(-1)^F$,
\ben \label{e3.4xxx}
&& Z_n(\beta, P,\mu,\omega=2\pi i/\beta) = Tr_P[e^{-\beta H + \beta\mu.Q } (-1)^F (2 h)^n]
\nonumber \\
&=& \sum_{Q}
 L^3 \int {d^3 k \over (2\pi)^3}  e^{- \beta {\vec k^2 \over 2 M_{BPS}(P,Q)}}\,
 Tr_{P,Q,\vec k=0} [e^{-\beta M_{BPS}(P,Q) 
 + \beta\mu.Q  } (-1)^F (2 h)^n]\nonumber \\
&\sim& \sum_Q e^{S_{BPS}(P,Q)-\beta M_{BPS}(P,Q)+\beta\mu.Q} L^3 (M_{BPS}/\beta)^{3/2}\, ,
\een
where $\sim$ denotes equality up to numerical factors and we shall choose the convention that
$L$, $\vec k$, $\beta$ and $M_{BPS}$ are all measured in the string units.
In writing the above equation we have imagined that the system is placed inside a box of size $L$
so that the density of states is of order $L^3 d^3 k / (2\pi)^3$.
We have used the non-relativistic formula for the kinetic energy since in the limit of large charges
the black hole is heavy. Relativistic corrections will be suppressed by additional powers of
charges in the denominator.

Defining 
\be \label{edefnubeta}
\nu\equiv \beta\,\mu \, , 
\ee
\refb{e3.4xxx} can be reversed as
\be\label{emaster}
e^{S_{BPS}(P,Q)} \sim L^{-3} (\beta/M_{BPS})^{3/2}
e^{\beta M_{BPS}(P,Q)} \int d^{n_v}\nu \, e^{-\nu .Q+\ln Z_n(\beta, P,\mu=\nu/\beta,\omega=2\pi i/\beta) }
\, ,
\ee
where $n_v$ is the total number of vector fields. 
The $\nu_i$ integrals in \refb{emaster} run along the imaginary axes so that \refb{emaster} gives
Fourier transform, but this will not play any role in our analysis since we shall analyze this using saddle point
approximation.
Our goal will be to compute the logarithmic
correction to $S_{BPS}$ by evaluating the right hand side of \refb{emaster}.

The integral on the right hand side receives contribution from the Euclidean rotating
black hole saddle point, satisfying the equation
\be
Q_i={\p \ln Z_n\over \p \nu_i}\, .
\ee
After carrying out the integration over $\nu_i$ we get, to the first subleading order, and ignoring
additive constants,
\ben \label{eindsub}
S_{BPS}(P,Q) &=& \beta M_{BPS}(P,Q)- \nu .Q+\ln Z_n(\beta, P,\nu/\beta,\omega=2\pi i/\beta)
\nonumber \\
&-& 3\, \ln\, L +{3\over 2} \ln{\beta\over M_{BPS}} -{1\over 2}  \ln \det {\p^2 \ln Z_n\over \p\nu_i\p\nu_j}\, ,
\een
evaluated at the saddle point.     
It will be understood from now on that all quantities must be evaluated at the saddle point. 
Even though the individual terms on the right hand side depend on $\beta$, $S_{BPS}$
computed from this formula is expected to be independent of $\beta$.

In the full geometry, the scalar moduli, including the string coupling, vary and as a result the 
scaling analysis that determines the dependence on $\lambda_Q/\lambda_P$, becomes
difficult. To avoid this problem we shall set the asymptotic moduli to be  their 
attractor values so that  the moduli remain frozen at their attractor values 
everywhere in the solution and we can apply the scaling analysis.
In fact in this case the solution just becomes a Euclidean Kerr-Newman black hole with
appropriate imaginary angular momentum so that we satisfy the $\beta\omega=2\pi i$
condition.\footnote{This can be seen as follows. 
In \cite{Chow:2014cca} the authors constructed charged rotating black
hole solutions in STU model and for special choice of charges these reduce to Kerr-Newman black holes
with constant scalars. Since STU model is a consistent truncation of $\NN=4$ and $\NN=8$ supergravity,
these black holes can also be regarded as solutions in $\NN=4$ and $\NN=8$ supergravity in four
dimensions for some special choice of charge vectors $Q,P$ and asymptotic values of the moduli fields.
We can now perform a continuous SL(2,R) S-duality transformation on the
moduli and charges to convert
this to a solution carrying arbitrary values of $P^2$, $Q^2$ and $Q.P$ preserving the combination
$Q^2P^2-(Q.P)^2$ and then a continuous SO(6,$r$) T-duality transformation to convert this to a solution
carrying arbitrary charge vectors $Q$ and $P$ preserving the combinations $Q^2$, $P^2$ and $Q.P$.
}
Explicit form of this solution can be found in \cite{H:2023qko}.
For the same reason, we shall
set $\beta$ to be of order $\lambda_P$ since the Schwarzschild radius of the black hole
solution measured in string units 
is of order $\lambda_P$. On the other hand the Schwarzschild radius is of
order $G_NM_{BPS}$ and in the string units the Newton's constant $G_N$ is of order $g_s^2
\sim \lambda_P/\lambda_Q$. This gives
\be
M_{BPS} \sim g_s^{-2}\lambda_P \sim \lambda_Q, \qquad {\beta\over M_{BPS}}\sim
{\lambda_P\over \lambda_Q}\, ,
\ee
and \refb{eindsub} may be written as
\ben \label{eindsubnew}
S_{BPS}(P,Q) &=& \beta M_{BPS}(P,Q)- \nu .Q+\ln Z_n(\beta, P,\nu/\beta,\omega=2\pi i/\beta)
\nonumber \\
&-& 3\, \ln\, L +{3\over 2} \ln{\lambda_P\over \lambda_Q} 
-{1\over 2} \ln \det {\p^2 \ln Z_n\over \p\nu_i\p\nu_j}\, .
\een
The dominant contribution to $Z_n$ comes from an euclidean rotating black hole saddle point
with parameters set by $P$, $\mu$, $\beta$ and $\omega=2\pi i/\beta$. 
While the usual euclidean rotating black hole saddle has complex metric, in this case the metric
is real due to imaginary $\omega$.
The logarithmic
contribution to $\ln Z_n$ comes from one loop determinant of massless field fluctuations around this
saddle point.
We shall now evaluate this separately for the $\NN=4$ theories and $\NN=8$ theories.

\subsection{$\NN=4$ theories}

We begin our analysis with the $\NN=4$ theories.
 We can divide the logarithmic contribution  to $\ln Z_n$
 into two parts, the non-zero mode contribution and the
zero mode contribution. The non-zero mode contribution is evaluated using the 
heat kernel expansion\cite{Vassilevich:2003xt}.
This consists of two steps. In the first step one includes both, the non-zero mode and the 
zero mode contribution, in the heat kernel.
One can compute this either by explicitly evaluating the eigenvalues of the kinetic operator or by
expressing the heat kernel in a general background as an integral of a local 
scalar constructed from
the background fields\cite{Vassilevich:2003xt}. 
Both approaches are supposed to give the same result. It was shown in 
\cite{Charles:2015eha,Karan:2019gyn} that 
for BPS black holes of the type we are analyzing, the local scalar to be integrated is 
proportional to the Euler density and hence its integral is a topological invariant, taking the same
value in the full Euclidean black hole geometry and the near horizon $AdS_2\times S^2$ geometry.
Hence we can use the result of the near horizon geometry computed in 
\cite{Banerjee:2010qc,Banerjee:2011jp}. It takes the form:
\be\label{eheat}
\left(- {169\over 45} + {124\over 45} \right) \ln a^2 = -2 \, \ln\, a\, ,
\ee
where $-169/45$ is the contribution from the
bosonic fields in the gravity multiplet and 124/45 is the contribution from the fermionic fields in the
gravitino multiplet. The matter multiplet contribution vanishes.
$a$ is the size of the black hole. 

The result given in \refb{eheat} is based on the understanding that each bosonic mode gives a contribution
of order $a$ to the partition function since the eigenvalues of the bosonic kinetic operator are
of order $a^{-2}$ and each fermionic mode gives a contribution of order $a^{-1/2}$
to the partition function since the eigenvalues of the fermionic kinetic operator are of order 
$a^{-1}$. However this is not true for the zero modes for which the eigenvalues vanish. Hence to 
get the actual contribution from the non-zero modes we need to remove the zero mode contribution.
These zero modes come from 
three translational and two rotational zero modes which are degrees of freedom of the metric and
$2n$ gravitino zero modes associated with the $2n$  supersymmetries that are broken by the
black hole solution. Note that rotation about the third axis leaves the black hole invariant and hence
does not generate a zero mode. Also note that rotation about the 1 and 2 axes will not be compatible
with the asymptotic boundary condition for generic angular momentum since the periodicity in Euclidean
time requires also a shift in the azimuthal angle and a rotation about 1 and 2 axes do not respect this
boundary condition. However for $\beta\omega=2\pi i$ the accompanying shift in the azimuthal angle
becomes $2\pi$ and hence the rotations about 1 and 2 axes generate allowed deformation of the
solution.  Together the 5 zero modes of the metric and $2n$ zero modes of the gravitini 
would give a contribution
\be \label{enumberzero}
5\ln a - n\, \ln\, a\, ,
\ee
if they are treated as non-zero modes. This contribution has 
been overcounted in \refb{eheat}. Subtracting this from \refb{eheat} we get the
non-zero mode contribution to the heat kernel:
\be\label{eheat2}
 -2 \, \ln\, a - 5\, \ln a + n\, \ln a = (n-7) \, \ln a\, .
\ee

We are however not quite done with the analysis of the 
non-zero mode contributions. Since our goal is to keep track of logarithms
of $\lambda_P$ and $\lambda_Q$ separately,  we need to determine whether $a$ in \refb{eheat2} is
$\lambda_Q$ or $\lambda_P$ or some sort of weighted geometric mean of these quantities. To fix
this we need to know which metric we should use to compute the size $a$ of the black hole
that appears in the non-zero mode contribution to the partition function. 
Since in weakly coupled string theory the string scale provides
the UV cut-off,  it is natural to assume that $a$ should be the size of the black hole measured in the
string scale, i.e.\ we should set $a=\lambda_P$. This can be stated formally by fixing the path integral
measure as follows. 
Let us suppose that
we have a rank $k$ covariant tensor field $C_{\mu_1\cdots \mu_k}$. Then we normalize the fields
$C_{\mu_1\cdots\mu_k}$ such that the quadratic part of the action has the form
\be \label{emeasure1}
\int d^4 x \, C_{\mu_1\cdots \mu_k} A^{\mu_1\cdots\mu_k; \nu_1\cdots \nu_k} 
C_{\nu_1\cdots \nu_k} \, ,
\ee
where $A^{\mu_1\cdots \mu_k;\nu_1\cdots \nu_k}$ is a second order differential operator that scales as
the inverse string metric and does not depend on the string coupling. 
Effectively this means that only the two derivatives contract with the inverse
metric; any factor of the inverse metric that might be
needed to contract the indices of $C_{\mu_1\cdots\mu_n}$ is absorbed into a redefinition of 
$C_{\mu_1\cdots\mu_n}$.
In other words we use tensors carrying tangent space indices which are contracted by flat metric.
Also the $\sqrt{\det G}$ factor and 
any function of the dilaton, e.g. $e^{-2\phi}$ that might have appeared as the coefficient of the
kinetic term, is removed by absorbing $(\det G)^{1/4}$ and the appropriate functions of $\phi$, e.g. $e^{-\phi}$, 
into a redefinition of the
tensor fields. With tensor fields normalized this way, we use flat measure for computing the
partition function
\be \label{emeasure2}
Z_n = \int \prod_{x,\mu_1,\cdots, \mu_n} dC_{\mu_1\cdots \mu_n}(x)\times \cdots \, e^\II\, (2h)^n \, ,
\ee
where $\cdots$ denote integration measure over other fields, determined in the same way and
$\II$ is the action.
A similar prescription holds for half integer spin fields where the kinetic term scales as the inverse
of a string frame 
vierbein that contracts with the single derivative in the kinetic term. Rest of the factors
involving the string metric and the dilaton are absorbed into the fields. The integration measure is 
again a
flat integration measure for fields normalized this way.
This procedure effectively sets the string scale
as the UV cut-off since the eigenvalues of the kinetic operator are determined by the size of the
black hole measured in string units.

We have verified in appendix \ref{sa} that this integration measure is compatible with the
integration measure in string field theory that follows from the BV formalism.

With this prescription the integration over each bosonic mode will generate a factor of the
size of the black hole {\it measured in the string metric} and 
 the integration over a pair of fermionic modes will generate a factor of the  inverse
 size of the black hole, also measured in the string metric. Therefore $a$ in \refb{eheat2} should
 be identified as $\lambda_P$ and we finally get the non-zero mode contribution to $\ln Z_n$ to be
\be\label{eheat3}
(n-7) \, \ln \lambda_P\, .
\ee

We now turn to the evaluation of the contribution from the zero modes. 
A zero mode integral involves integration over certain modes of the fields, with integration 
measure normalized as in \refb{emeasure1}, \refb{emeasure2}. For the zero modes of the metric, it is
easy to see that the fluctuations $h_{\mu\nu}$, 
normalized as in \refb{emeasure1}, \refb{emeasure2}, appears
in the expression for the string metric $G_{\mu\nu}$ as
\be \label{emetricexpand}
G_{\mu\nu} = G^B_{\mu\nu} + g_s\, h_{\mu\nu}\, ,
\ee
where $G^B_{\mu\nu}$ is the background value of the string metric. With this the quadratic term
involving $h_{\mu\nu}$ will have a factor of $g_s^2$ that cancels the background value of $e^{-2\phi}$
that appears in the action. On the other hand the size dependence from the $\sqrt{\det G}$ factor
cancels the two factors of $G^{\mu\nu}$ that are needed to be contracted with the two factors of 
$h_{\mu\nu}$, and we are left with a single factor of $G^{\mu\nu}$ that is to be contracted with the two
derivatives. This agrees with the general prescription given in \refb{emeasure1}, \refb{emeasure2}.

To evaluate the contribution from the metric zero modes we have to find the range of integration over
these zero modes. For these we note that the zero modes are associated with diffeomorphism
transformations with parameters that do not vanish at infinity. Since under a diffeomorphism with
parameter $\xi^\mu$, 
$\delta G_{\mu\nu}=D_\mu \xi_\nu + D_\nu\xi_\mu$, we can write
\be
\delta h_{\mu\nu} = g_s^{-1}\left[\p_\mu (G^B_{\nu\rho} \xi^\rho) + \p_\nu (G^B_{\mu\rho} \xi^\rho) \right]
+ \hbox{non-linear terms}\, .
\ee
Noting that $G^B_{\mu\nu}\sim\lambda_P^2$, we see that if the range of $\xi^\rho$ is $\Delta\xi$, then
the range of the corresponding zero mode of the metric scales as 
\be\label{erangex}
g_s^{-1} \, \lambda_P^2 \, \Delta\xi\, .
\ee

Let us first consider the translation zero modes. They must span a physical distance equal to the
length $L$ of the box in which we have enclosed the system. Since $L$ is measured in the string
metric that scales as $\lambda_P^2$, the coordinate distance associated with this length is 
$L/\lambda_P$.\footnote{Since in our convention the string metric scales as $\lambda_P^2$, the
asymptotic metric is also of order $\lambda_P^2$.}
\refb{erangex} now shows that the three translation modes give a net factor of
\be\label{etrans}
\left( g_s^{-1} \lambda_P^2 {L \over \lambda_P}\right)^3\sim g_s^{-3} L^3 \lambda_P^3 \sim 
L^3 \lambda_P^{3/2} \, \lambda_Q^{3/2} \, ,
\ee 
where in the last step we used $g_s^2 =\lambda_P/\lambda_Q$.

For rotational zero modes 
the analysis is similar but since the parameter is the angle of rotation it has a finite range.
Hence using \refb{erangex} we get the following net contribution from the rotational zero modes
\be \label{erotcont}
\left( g_s^{-1} \lambda_P^2 \right)^2 \sim g_s^{-2} \lambda_P^4 \sim \lambda_P^3 \lambda_Q\, .
\ee

Next we turn to the gravitino zero modes. Let us denote by $\psi_\mu$ the 
gravitino zero mode normalized
according to the procedure described above. Then the only dependence of the gravitino kinetic
term on $\lambda_P$ or $\lambda_Q$ comes through the inverse vierbein contracted with a single
derivative appearing in the kinetic term. The inverse vierbein scales as $\lambda_P^{-1}$. The
rotation generator $J_3$, constructed from the action using the Noether procedure, will involve replacing
the derivative by the spin of the zero mode and hence will scale as $\lambda_P^{-1}$. Therefore 
the $(2h)^n$ factor will produce a factor of $\lambda_P^{-n}$ times the product of  $2n$ zero modes. 
The zero mode integral now takes the form
\be \label{egravitino}
\int \prod_{i=1}^{2n} d\psi^i_0 \, \lambda_P^{-n} \prod_{i=1}^{2n} \psi^i_0 \sim  \lambda_P^{-n}\, .
\ee
There are also non-zero mode contributions to $2h$ in the form:
\be
2h = \hbox{zero mode contribution + non-zero mode contribution}\, .
\ee
When we raise it to $n$-th power and express the result in a binomial expansion, any term
involving non-zero mode contribution will have less than $n$ powers of the zero mode contribution.
Hence it will have less than $2n$ factors of $\psi^i_0$ and
the contributions will vanish after
integration over the zero modes  in
\refb{egravitino}.

Adding the logarithm of  \refb{etrans}, \refb{erotcont} and \refb{egravitino} 
we get the net zero mode contribution to $\ln Z_n$:
\be \label{enetzero}
3\ln L + {3\over 2} \ln\lambda_P  + {3\over 2} \ln\lambda_Q + 3\ln\lambda_P
+\ln\lambda_Q - n\ln \lambda_P =3\ln L + \left({9\over 2}-n\right) 
\ln\lambda_P+ {5\over 2} \ln\lambda_Q\, .
\ee
Adding it
to the non-zero mode contribution \refb{eheat3}, 
we get the net logarithmic term in $\ln Z_n$:
\be\label{etothet}
(n-7)\ln\lambda_P + 3\ln L + \left({9\over 2}-n\right) 
\ln\lambda_P+ {5\over 2} \ln\lambda_Q = 3\ln L - {5\over 2} \ln\lambda_P + {5\over 2} \ln \lambda_Q\, .
\ee

Our next task is to estimate $\det {\p^2 \ln Z_n\over \p\nu_i\p\nu_j}$. 
At the saddle point $\ln Z_n$ scales as $S_{BH}\sim \lambda_P\lambda_Q$. Since 
$Q_i=\p\ln Z_n/\p\nu_i$ scales as 
$\lambda_Q$, we must have, at the saddle point
\be
\nu_i \sim \lambda_P\, .
\ee
Hence, taking into account the fact that $\nu_i$ is an $r+6$ dimensional vector, we get
\be\label{eldetjac}
\left(\det \, {\p^2\ln Z_n\over \p\nu_i\p\nu_j}\right)\sim (\lambda_Q/\lambda_P)^{(r+6)}\, .
\ee

Substituting \refb{etothet} and \refb{eldetjac} into \refb{eindsubnew} we get
\ben \label{eindsubfin}
S_{BPS}(P,Q) &=& \beta M_{BPS}(P,Q)- \nu .Q+\ln Z_{\rm classical}
\nonumber \\
&+& 3\ln L - {5\over 2} \ln\lambda_P + {5\over 2} \ln \lambda_Q -
3\, \ln\, L +{3\over 2} \ln{\lambda_P\over \lambda_Q} 
-{r+6\over 2} \ln{\lambda_Q\over \lambda_P}\nonumber \\
&=& S_{\rm classical} -{r+4\over 2} \ln {\lambda_Q\over \lambda_P}\, ,
\een
where $S_{\rm classical}$ is the classical entropy given by the terms in the first line.
The coefficient of the $\ln(\lambda_Q/\lambda_P)$ term differs from the microscopic prediction 
\refb{e2}, \refb{e3} by 1. We shall resolve this discrepancy in section \ref{s4}.

\subsection{$\NN=8$ theory}

We now turn to type II string theory on $T^6$. We first consider the contribution from the
non-zero modes. 
The heat kernel contribution, that includes the zero modes, was found in \cite{Banerjee:2011jp} to be
\be \label{eheatii}
\left(- {169\over 45} + {124\over 45}  - {136\over 45} - {44\over 45} \right) \ln a^2 = -10 \, \ln\, a\, ,
\ee
where the first two terms are the contributions in the $\NN=4$ theories
and the last two
are the contributions from the extra bosonic and fermionic modes that arise in type II on 
$T^6\times R^{3,1}$. The original analysis in \cite{Banerjee:2011jp}  was carried out in the
near horizon $AdS_2\times S^2$ background, but following the same logic as in the case of
$\NN=4$ theories, one can argue that
\refb{eheatii} is also valid for the full black hole solution with $\beta\omega=2\pi i$.

The zero mode contribution to the heat kernel is given by 
the same expression as \refb{enumberzero} since the zero mode structure is the same except that $n$
is different. Subtracting this from \refb{eheatii}, and replacing $a$ by $\lambda_P$ using the same logic
as in the case of $\NN=4$ theories, we get the net non-zero mode contribution to $\ln Z_n$
to be
\be\label{enz1ii}
(n-15)\, \ln\lambda_P\, .
\ee

The analysis of the zero mode contribution to $Z_n$  is also identical. We get the same result as
\refb{enetzero}:
\be \label{ezeroii}
3\ln L + \left({9\over 2}-n\right) 
\ln\lambda_P+ {5\over 2} \ln\lambda_Q\, .
\ee
Adding this to \refb{enz1ii} we get the net logarithmic correction to $\ln Z_n$:
\be\label{exx1}
3\ln L -{21\over 2} \ln\lambda_P + {5\over 2} \ln\lambda_Q\, .
\ee

Finally, let us estimate $\det {\p^2 \ln Z_n\over \p\nu_i\p\nu_j}$.  When the chemical potential
associated with the RR sector gauge fields are set to zero, the matrix ${\p^2 \ln Z_n\over \p\nu_i\p\nu_j}$
takes a block diagonal form with no mixing between the RR and NSNS sector fields. Hence we
can evaluate the determinants of the two blocks separately and take their product.
First let us consider the NSNS sector block.
As before, at 
the saddle point $\ln Z_n$ scales as $S_{BH}\sim \lambda_P\lambda_Q$. 
The $Q_i$'s coming from NSNS sector charges scale as 
$\lambda_Q$. Hence for these we must have, at the saddle point
\be
\nu_i \sim \lambda_P\, .
\ee
On the other hand for the RR sector charges there is no intrinsic difference between
electric and magnetic charges since they can be transformed into each other by T-duality
transformation. Hence the $Q_i$'s scale as $\sqrt{\lambda_Q\lambda_P}$ and
\be
\nu_i \sim \sqrt{\lambda_Q\lambda_P}\, .
\ee
There is another way to see this scaling of $\nu_i$. 
For this let us work in the coordinate system where $\lambda_P^2$ appears as the overall factor in
the metric, and the compact 
coordinate ranges (including the period of the euclidean time coordinate) are of
order unity, the horizon is at $r\sim 1$ and the asymptotic region corresponds to $r>>1$. 
In this coordinate system the asymptotic value of the gauge fields $A^{(i)}_\tau$
will be order
$\nu_i$ since $\int A^{(i)}_\tau d\tau$, which is invariant under coordinate transformation, is of order
$\nu_i$. Since $A^{(i)}_\tau$ vanishes at the horizon and hence changes by $\nu_i$
over a distance scale of order unity,
the field strengths are of order $\nu_i$. When we square this and contract with two
factors of the inverse metric to get the Lagrangian density, it scales as $\lambda_P^{-4}$.
Since the kinetic term of the gauge fields arising from
the RR sector does not carry the $e^{-2\phi}$ factor that the kinetic term of NSNS sector gauge
fields carries\cite{Polchinski:1998rr}, we do not have any other factor in the Lagrangian density.
We now multiply this by $\sqrt {\det G}\sim \lambda_P^4$ and integrate over the space-time
volume to get the action. This gives the contribution to $\ln Z_n$ to be of order
\be
\lambda_P^4 \times \lambda_P^{-4}\times \nu_i^2 \sim \nu_i^2\, .
\ee
Since $\ln Z_n$ scales as $\lambda_Q\lambda_P$, we assign the scale $\sqrt{\lambda_Q\lambda_P}$ to
the $\nu_i$'s associated to the RR sector gauge fields so that $\p^2\ln Z_n/\p\nu_i\p\nu_j\sim 1$. 
In contrast, due to the explicit factor of $e^{-2\phi}$ in the
kinetic term of the NSNS sector fields, $\ln Z_n$ will get contribution of order 
$g_s^{-2}\nu_i^2$ for the boundary
values $\nu_i$ 
of the NSNS sector gauge fields. Since $g_s^2$ scales as $\lambda_P/\lambda_Q$, we
see that $\p^2 \ln Z_n/\p\nu_i\p\nu_j\sim g_s^{-2}\sim \lambda_Q/\lambda_P$,
and the corresponding $\nu_i$'s must scale as $\lambda_P$.

Taking into account the fact that we have 12 NSNS sector gauge fields and 16 RR sector
gauge fields\footnote{Note that we had restricted the charge vector to lie entirely
in the NSNS sector so that the near horizon value of the string coupling would be small.
This statement would then amount to evaluating the gravitational path integral at zero 
chemical potential for the RR sector charges. As we have shown, the change of ensemble 
from grand canonical to microcanonical produces factors of order unity for the RR sector 
charges.}, we get
\be \label{exx2}
\left(\det \, {\p^2\ln Z_n\over \p\nu_i\p\nu_j}\right)\sim (\lambda_Q/\lambda_P)^{12}\, .
\ee

Substituting \refb{exx1} and \refb{exx2} into \refb{eindsubnew}
\ben \label{eindsubiifin}
S_{BPS}(P,Q) &=& \beta M_{BPS}(P,Q)- \nu .Q+\ln Z_{\rm classical}(\beta, P,\nu/\beta,\omega=2\pi i/\beta)
\nonumber \\
&+&3\ln L -{21\over 2} \ln\lambda_P + {5\over 2} \ln\lambda_Q  -3\, \ln\, L +{3\over 2} \ln{\lambda_P\over \lambda_Q} 
-6 \, \ln{\lambda_Q\over \lambda_P} \nonumber \\
&=& S_{\rm classical}  - 8 \ln \lambda_P - 5 \ln {\lambda_Q\over
\lambda_P}\, .
\een
Again we see that the coefficient of the $\ln(\lambda_Q/\lambda_P)$ term 
differs from the microscopic prediction 
\refb{e2}, \refb{e4} by 1. We shall resolve this discrepancy in section \ref{s4}.

\sectiono{Axion to Kalb-Ramond field} \label{s4}

Let us review the logic behind our analysis so far. We have chosen the charges and the
asymptotic values of the moduli fields so that the string coupling remains small everywhere in the
Euclidean black hole saddle point. In this case the dynamics of the fluctuating fields in this
background should be described by weakly coupled string theory and we should be able to
carry out our analysis using the string scale as the UV cut-off.

While the logic is sound, there is one caveat.  
In our analysis we have dualized the Kalb-Ramond 2-form 
field  into the axion field forming the
real part of the complex modulus $\tau$, whose imaginary part is the inverse of the square of the
string coupling. However 
the path integral in string field theory, which should be the valid description at weak coupling, is over
the Kalb-Ramond field instead of the axion.
So we need to go back to the description in terms of the 2-form field $B_{\mu\nu}$. 
It was shown in \cite{Duff:1980qv} that the difference in the value of the heat kernel under
a duality transformation comes entirely from the zero modes. Therefore we need to remove
the contribution from the zero modes of the axion field from the results in section \ref{s2} and
include the effect of integration over the zero modes of the 2-form field. The axion field, being a scalar,
has no zero modes, since the only possible zero mode is a constant and this is not normalizable.
So we do not have to remove anything from the results of section \ref{s2}. On the other hand, 2-form
field $B_{\mu\nu}$ does have normalizable zero modes in the Euclidean Kerr-Newman black
holes of the type we are considering. They are simply the electromagnetic field associated with
the electrically charged euclidean Kerr-Newman black hole and its Hodge dual,  which would be the
electromagnetic field associated with the magnetically charged euclidean Kerr-Newman black 
hole.\footnote{This is consistent with the fact that the black hole space-time has Euler number 2,
and there are no normalizable zero and one forms and hence also no normalizable three and four forms.
Due to the absence of normalizable zero and one forms there are also no ghost zero modes.}
The normalizability of these modes can be seen by noting that in cartesian coordinates these modes
decay as $1/r^2$ for large radial distance $r$. Hence the square of these forms fall off as $1/r^4$ which
integrates to finite value, since the spatial integrals generate a factor of $r^2 dr$ and the time integral
is finite due to finite size of the time circle.

We shall now evaluate the contribution from these two zero modes.
Since $B_{\mu\nu}$ is a rank two tensor, the analysis begins as in the case of the metric.
The analog of \refb{emetricexpand} takes the form:
\be\label{eBexpand}
B_{\mu\nu} = B^B_{\mu\nu} + g_s\, b_{\mu\nu}\, ,
\ee
where $B^B_{\mu\nu}$ is the background value of the 2-form field which vanishes for Kerr-Newman
black hole.
To find the range of integration over the zero modes of
$b_{\mu\nu}$, we note that the euclidean world-sheet of the fundamental string, wrapped
on some 2-cycle ${\cal C}$, has a term in its action of the form $i\int_{\cal C} B$. Since this
appear in the exponent, $\int_{\cal C} B$ has a period of order unity. For the euclidean black hole
the relevant 2-cycles can be taken to be the celestial sphere and the $r$-$\tau$ plane.\footnote{Note
that even though the $r$-$\tau$ plane is non-compact, the integral of the zero mode 
wave-function along
this plane is finite. This is sufficient to establish the periodicity of the zero mode.}
Since  $\int_{\cal C} B$ for these cycles do not have any dependence on the metric or the
dilaton,  the integration range 
of the zero modes of $B_{\mu\nu}$ will be of order unity. 
This in turn means that the integration range of the zero modes of $b_{\mu\nu}$ 
is of order $g_s^{-1}$ and
the integration over the two zero modes produces a factor of $g_s^{-2} \sim \lambda_Q/
\lambda_P$. This gives an extra
logarithmic correction to the entropy of the form
\be\label{eextraterm}
\ln {\lambda_Q\over \lambda_P}\, .
\ee
Adding this to \refb{eindsubfin} and \refb{eindsubiifin} we get the final results for the logarithmic terms in
$\NN=4$  and $\NN=8$ theories:
\ben
S_{BPS}(P,Q) &=& S_{\rm classical}  - {r+2\over 2} \ln {\lambda_Q\over
\lambda_P}, \qquad \hbox{for $\NN=4$ theories}, \nonumber \\
S_{BPS}(P,Q) &=& S_{\rm classical}  - 8 \ln \lambda_P - 4 \ln {\lambda_Q\over
\lambda_P}, \qquad \hbox{for $\NN=8$ theory}\, .
\een
This agrees with the microscopic results given in \refb{e2}, \refb{e3} and \refb{e4}.

One might worry that in type II theory we also have 2-form fields $B'_{\mu\nu}$
coming from the RR sector. The sources of these fields are D-strings or D-branes 
with all but one spatial world-volume
directions wrapped along some compact direction, and one can repeat the analysis.
However,  the fields $B'_{\mu\nu}$ that appear in the D-brane world-volume action
as $\int_{\cal C} B'$ have no factor of $e^{-2\phi}$ in their space-time action\cite{Polchinski:1998rr} 
and hence 
the $g_s$ factor will be missing in the analog of \refb{eBexpand}. 
Therefore integration over these zero modes does not produce any factor of $g_s$.

\sectiono{$\NN=2$ theories}

In this section we shall generalize the analysis to half-BPS black holes in
heterotic string theory on $K3\times T^2$, as well as
various orbifolds of heterotic string theory on $T^6$
that preserve
$\NN=2$ supersymmetry, {\it e.g.} the one described in \cite{Ferrara:1995yx}. 
If the resulting theory has $n_V$ vector multiplet and $n_H$ hyper-multiplet
fields, then the net contribution to the logarithmic correction to the entropy due to the heat kernel
is given by the expression\cite{Sen:2012kpz}
\be
-8\ln a \left({53\over 90}-{589\over 720}\right) +{1\over 6}\,\ln a\, (n_H-n_V) = {1\over 6} \, (11 - n_V + n_H)
\, \ln a\, ,
\ee
where $53/90$ represents the contribution to the heat kernel from the bosonic fields of the gravity
multiplet and $-589/720$ represents the contribution to the heat kernel from the fermionic fields 
of the gravity
multiplet. This is the analog of \refb{eheat}. Subtracting the zero mode contribution \refb{enumberzero}
from this, and replacing $a$ by $\lambda_P$ as in the case of $\NN=4,8$ theories described 
earlier, we get the net contribution to the entropy from the non-zero modes to be
\be\label{en2}
{1\over 6} \, (11 - n_V + n_H)
\, \ln \lambda_P - (5-n)\ln\, \lambda_P\, .
\ee
In this case $n=2$ since a half BPS black hole in $\NN=2$ supersymmetric theory breaks four
supersymmetries. 

The zero mode contribution is given by \refb{ezeroii}
\be \label{ezeron2}
3\ln L + \left({9\over 2}-n\right) 
\ln\lambda_P+ {5\over 2} \ln\lambda_Q\, .
\ee
with $n=2$. Adding this to \refb{en2} we get the net logarithmic
contribution to $\ln Z_n$ to be:
\be
{1\over 6} \, (8 - n_V + n_H)
\, \ln \lambda_P +3\ln L + {5\over 2} \ln\lambda_Q\, .
\ee
Also, we have $n_V+1$ gauge fields, -- $n_V$ from the vector multiplets and one from the
gravity multiplet, all coming from the NS sector. Hence we have the
analog of \refb{eldetjac}
\be\label{eldetjacn2}
\left(\det \, {\p^2\ln Z_n\over \p\nu_i\p\nu_j}\right)\sim (\lambda_Q/\lambda_P)^{n_V+1}\, .
\ee
Finally changing the integration variables from the axion to the Kalb-Ramond field produces
an extra term in the entropy
\be
\ln{\lambda_Q\over \lambda_P}\, ,
\ee
as in \refb{eextraterm}. 
Substituting these into \refb{eindsubnew} we get
\be
S_{BPS} = S_{\rm classical} + {1\over 6} \, (23 - n_V + n_H)
\, \ln \lambda_P +  \left(2 - {n_V+1\over 2}
\right)\, \ln{\lambda_Q\over \lambda_P}\, .
\ee
At present there are no microscopic results that one can compare this with.

\bigskip

\noindent{\bf Acknowledgement:} 
This work was supported by 
Department of Atomic Energy, Government of India, under project no.~RTI4019.
The work of A.S. was supported by the ICTS-Infosys Madhava 
Chair Professorship. We acknowledge ChatGPT for providing useful information.

\appendix

\sectiono{Measure from string field theory} \label{sa}

In our analysis the measure that we have used 
was described between \refb{emeasure1} and \refb{emeasure2}. 
The goal of this section will be to show that this is the measure that follows from string field theory.

In string field theory
the BV formalism fixes the measure. 
In particular, the integration measure over the string field is flat. (See {\it e.g.}
\cite{Sen:2024nfd} for a review.)
If this has to agree
with the integration measure described in \refb{emeasure1}, \refb{emeasure2}, then
the kinetic term of the string field should have the property that under a change in the background
string metric $G_{\mu\nu}$ or the dilaton $\phi$, 
the only change in the kinetic term will be induced by the components of
$\delta G^{\mu\nu}$ contracted with the derivatives acting on the fields. We shall verify this using
the example of  bosonic 
string theory in locally flat space-time but the result is quite general. Let us suppose
that the tensor field $C$ is the metric fluctuation $h_{\mu\nu}$
itself, and we want to study the effect of deforming the
background string frame
metric. Then the change in the quadratic term in the action will be proportional to the
three point function
\be \label{ethree}
\int d^D k\,  \delta G_{\mu\nu} \, h_{\rho\sigma}(k)\, h_{\alpha\beta}(-k)\, 
\left\langle c \bar c\,  \p X^\mu \bar\p X^\nu (0) \,   c\bar c \, \p X^\rho \bar\p X^\sigma e^{ik.X}  (1)
\,   c\bar c \, \p X^\alpha \bar\p X^\beta  e^{-ik.X}(\infty)
\right\rangle \, .
\ee
We shall assume that $h_{\mu\nu}(k)$ satisfies $k^\mu h_{\mu\nu}=0=k^\nu h_{\mu\nu}$ for
simplicity, since the vertex operator associated with $h_{\mu\nu}$ not satisfying these conditions
have additional terms\cite{Sen:2024nfd}. 
In this case the three point function appearing in \refb{ethree} gets contribution
only from terms where $\p X^\mu$ and $\bar\p X^\nu$ factors inside the correlation function contract with
$e^{\pm ik.X}$, bringing down powers of momentum. There are no contributions where
$\p X^\mu$ is contracted with $\p X^\rho$ or $\p X^\alpha$, or where $\bar\p X^\nu$ is contracted 
with $\bar\p X^\sigma$ or $\bar\p X^\beta$. This shows that the change in $G_{\mu\nu}$ only affects
the kinetic term through contraction with momenta and not through contraction with the indices of the
tensor fields $h_{\rho\sigma}$ or $h_{\alpha\beta}$. This analysis easily generalizes to other tensor fields,
the main thrust of the argument being that the required correlation function has odd number of $\p X$
and odd number of $\bar\p X$ insertions, and hence the only way to get a non-zero answer will be
to contract a $\p X$ and a $\bar\p X$ with the $e^{ik.X}$ factors.

We also need to verify that the kinetic term does not change under a change in the dilaton field.
For this we note that the zero momentum dilaton vertex operator is proportional to
$(c\p^2 c - \bar c \bar\p^2 \bar c)$. So under a change in the dilaton proportional to
$\delta\phi$, the kinetic term for metric fluctuation will change by an amount proportional to
\be \label{ethreedil}
\int d^D k\,   \delta \phi \, h_{\rho\sigma}(k)\, h_{\alpha\beta}(-k)\, 
\left\langle  (c\p^2 c - \bar c \bar\p^2 \bar c) (0) \,   c\bar c \, \p X^\rho \bar\p X^\sigma e^{ik.X}  (1)
\,   c\bar c \, \p X^\alpha \bar\p X^\beta  e^{-ik.X}(\infty)
\right\rangle \, .
\ee
This vanishes by ghost number conservation. This proves the desired result.

This analysis can be easily generalized to
the NSNS sector fields in type II string theory and NS sector states in superstring theory. 
For RNS, NSR and RR sector fields in type II string theory and R sector states in the heterotic string
theory,  the vanishing of the zero momentum dilaton coupling follows in the same way 
using ghost number conservation, but the effect of background metric requires
a more detailed analysis. 
For R sector states we need to show that 
the dependence of
the kinetic term on the metric comes through the contraction of the single $\p_\mu$ factor to the inverse
vierbein, but as discussed in \cite{Sen:2015nph}, in string field theory the local Lorentz symmetry is already gauge fixed
and the vierbein appears as the symmetric square root of the metric. 
The desired result can be established following \cite{Mamade:2025jbs}
but we shall not give the details of the analysis
here.

\end{document}